\documentclass[twocolumn]{aastex631}

\usepackage{amsmath}
\usepackage{graphicx}
\usepackage{bm}
\usepackage{tabularx}
\usepackage{dcolumn}
\usepackage{overpic}
\usepackage{changepage}
\usepackage[version=4]{mhchem}
\usepackage{xcolor}
\usepackage{multirow}
\usepackage{makecell}

\def\simge{\mathrel{\rlap{\raise 0.511ex
       \hbox{$>$}}{\lower 0.511ex \hbox{$\sim$}}}}
\def\simle{\mathrel{\rlap{\raise 0.511ex
        \hbox{$<$}}{\lower 0.511ex \hbox{$\sim$}}}}

\makeatletter
\@ifundefined{citep}{\newcommand{\citep}[1]{\cite{#1}}}{}
\@ifundefined{citet}{\newcommand{\citet}[1]{\cite{#1}}}{}
\makeatother

\usepackage{xcolor}
\usepackage{soul}
\usepackage{cancel}

\setstcolor{red}

\newcommand{\redout}[1]{}

\shorttitle{Mass-Radius-Space Priors for the Neutron Star EOS}
\shortauthors{Sun, Zhao, \& Lattimer}

\begin{document}

\title{A New Bayesian Framework with Mass-Radius-Space Priors to Constrain the Neutron Star Equation of State}

\correspondingauthor{Tianqi Zhao}

\author[0009-0005-9688-4369]{Boyang Sun}
\affiliation{Department of Physics \& Astronomy, Stony Brook University, Stony Brook, NY 11794, USA}

\author[0000-0003-4704-0109]{Tianqi Zhao}
\affiliation{Institute for Nuclear Theory, University of Washington, Seattle, WA 98195, USA}
\affiliation{N3AS, University of California, Berkeley, Berkeley, CA 94720, USA}

\author[0000-0002-5907-4552]{James M. Lattimer}
\affiliation{Department of Physics \& Astronomy, Stony Brook University, Stony Brook, NY 11794, USA}

\begin{abstract}
We propose a new Bayesian framework to infer the neutron star equation of state (EOS) from mass and radius observations and neutron matter theory by defining priors that directly parameterize mass-radius space instead of pressure-energy density space.  We use direct and accurate inversion approximations to map mass-radius relations to the underlying EOS.  We systematically compare its EOS inferences with those inferred from traditional EOS parameterizations, taking care to quantify the systematic prior uncertainties of both. 
Our results show that prior uncertainties should be included in all Bayesian approaches. The alternative framework provides broader and more uniform coverage of the physically allowed mass-radius space, especially small radius configurations, with enhanced computational efficiency. Substantially reduced dependence on prior choices seems to result from having both prior and posterior distributions in the same observational mass-radius space. 
Results indicate that direct parameterization in observed space offers a robust and efficient alternative to traditional methods.
\end{abstract}


\section{Introduction}
The equation of state (EOS) of dense matter at densities in excess of the nuclear saturation density ($n_s\simeq0.16$ fm$^{-3}$), such as is found in neutron stars \citep{Lattimer2001,PhysRevC.109.055801}, is not easily accessible from terrestrial experiments or first-principle theoretical calculations.  But since the EOS, specifically the pressure-energy density relation for cold matter, is all that is needed to predict neutron star observables such as masses, radii, moments of inertia, tidal deformabilities and oscillation frequencies, observations can constrain it.  
Due to the lack of an exact analytical method to invert Tolman-Oppenheimer-Volkoff (TOV) structure equations applicable to neutron stars, Bayesian approaches have been extensively used to infer limits to the equations of state from observations of neutron stars and nuclear experiments and theory \cite{Brandes2023, Rutherford2024,kalita2025observable, dong2025}. In the traditionally-used framework, the model prior probability ${\cal P}({\cal M})$ is associated with selected parameterized EOS.  
Each EOS model, together with each possible set of parameters, generates a particular model ${\cal M}(\theta, \mathcal{E})$ in terms of the EOS parameters $\theta$ and central energy densities $\mathcal{E}$.  Another input is the prior probability associated with the observed data, usually probability distributions in mass-radius ($M{\mbox-}R$) space, which we call ${\cal P}({\cal D})$. The posterior distribution is then
\[
{\cal P}({\cal M}(\theta, \mathcal{E})|{\cal D})\propto{\cal P}({\cal D}|{\cal M}(M,R)){\cal P}({\cal M}(\theta, \mathcal{E})),
\]
which inevitably involves a mapping $(\theta, \mathcal{E})\rightarrow(M,R)$ to obtain the likelihood ${\cal P}({\cal D}|{\cal M}(M,R))$ from the sampling of prior distributions ${\cal P}({\cal M}(\theta, \mathcal{E}))$.
Moreover, the posterior distributions will somewhat depend on the prior choices. 
Even when only considering a single EOS parameterization, one also has to consider how its parameters are chosen. For example, should one choose  parameters uniformly within their physically allowed bounds, or should their parameters be chosen uniformly in log space? Should parameters be chosen so that central values are emphasized, or should they be chosen to uniformly cover the inferred mass-radius ($M$-$R$) space? 

These arbitrary choices  contribute to the inherent prior uncertainties that add to observational uncertainties in estimating the underlying EOS.  In most analyses to date, these prior uncertainties are largely unquantified, let alone considered, but could be comparable to the observational or theoretical uncertainties. Some recent works have, however, studied how the choices of EOS parametrization or astrophysical assumptions could affect the Bayesian inferences \citep{Li2025, kedia2025}.  We aim to further clarify prior uncertainties and attempt to quantify them.

We suggest an alternative approach in which $M$-$R$ space is  directly parameterized, which naturally circumvents the issue of EOS parameterizations. In this new framework, both the prior and posterior distributions are in $M$-$R$ space,  which may form a more consistent prior. 
After applying constraints from astronomical observations, causality, a lower limit for the neutron star maximum mass ($M_{\rm max}$), and neutron matter theory \citep{Drischler_2021}, it is possible to obtain the posterior distributions of EOS variables, i.e., pressure, energy density, chemical potential and number density, using $M$-$R$ inversion formulae developed by \citet{Sun2025}. Although the posteriors of this alternative method still depend on choices of modeling priors, these are now defined in $M$-$R$ space instead of EOS space. Furthermore, $M$-$R$ curves can be chosen to ensure that the distribution of radius values at a typical mass, such as $M = 1.4\,M_\odot$, are relatively uniform.

In order to estimate systematic prior uncertainties for both new and traditional frameworks, we compare the posteriors from a series of calculations employing different EOS or $M$-$R$ space parameterizations, as well as different theoretical neutron matter EOS choices.

\section{Methodology}
The alternative method we propose, which we refer to as the Parameterized Mass-Radius (PMR) method, is practical because we employ a highly accurate analytical scheme to invert the TOV equations [$(M,R)\rightarrow(P, \mathcal{E})$]. Specifically, given an entire $M$-$R$ curve, one determines the masses and radii at 11 fractional maximum mass ($M_{\rm {max}}$) points $f\in{[1,0.95,0.90,0.85,0.8,0.75,2/3, 0.6, 0.5, 0.4,1/3]}$. The thermodynamic properties at the center of a star of mass $M_f$ can be approximated using Eq.~(22) of \cite{Sun2025} as
\begin{eqnarray}
    G_{f}&\simeq&a_{Gf}\left({M_{max}\over M_\odot}\right)^{b_{Gf}}\left({R_{g}\over10{\rm~km}}\right)^{c_{Gf}}\left({R_{h}\over10{\rm~km}}\right)^{d_{Gf}},~~~~
\label{eq:eqfit1}\end{eqnarray}
where $G_f$ represents the central energy density $\mathcal{E}$, pressure $P$, sound speed $c_s/c =\sqrt{dP/d{\mathcal{E}}}$, chemical potential $\mu$, or baryon number density $n$ of a star with mass $M_f=M/M_{\rm max}$.  $a_{Gf}$, $b_{Gf}$, $c_{Gf}$, and $d_{Gf}$ are constant fitting parameters.   $R_g$ and $R_h$ are the radii of stars at fractional maximum masses $M_g$ and $M_h$, where $g,h\in[f]$; the values of $g$ and $h$ are also parameters of the scheme. Therefore, the scheme, for each thermodynamic quantity, has a $6\times 11$ matrix consisting of 6 constants ($a_{Gf}, b_{Gf}, c_{Gf}, d_{Gf}, g, h$) for each $f$. The central conditions for masses intermediate to gridded values at $M_f$ are determined by logarithmic interpolation.  The accuracies of this scheme, for configurations without first-order phase transitions, are generally better than 0.5\% \citep{Sun2025}.

There is no unique way to parameterize $M$-$R$ curves. In the following, we describe the specific PMR construction adopted in this work, while its mathematical definition is given in Appendix~\ref{ap:PMR}.

\subsection{Establishing the $M$-$R$ Boundary\label{sec:MR_boundary}}
The first step in the PMR scheme is to establish the physically-realistic boundaries of $M$-$R$ space, as investigated, for example, in \citet{Drischler_2021}. This step is comparable to traditional approaches used to define the boundaries of the allowed EOS in baryon density–chemical potential space prior to EOS parameterization \citep{zhou2025reexamining,gorda2025constrained}.  For the low-density regime we adopted the SLy4 crustal EOS \citep{chabanat1998skyrme} below a density of 0.04 fm$^{-3}$, represented as a piecewise polytropic (PP) \citep{Zhao2022}.  Above this density, up to a transition density $n_t=1.5n_s$, we use the central value together with the 1-$\sigma$ uncertainty band from the N$^3$LO $\chi$EFT calculations of neutron star matter \citep{Drischler_2021,buqeye_result,N3LO}.  At the density $n_t$, for the central values of the N$^3$LO EOS, the NS mass is $M_t\simeq0.45M_\odot$, the NS radius is $R_t\simeq13.0$ km, and the pressure is $P_t\simeq9$ MeV fm$^{-3}$.

\begin{figure*}
    \includegraphics[width=0.495\linewidth]{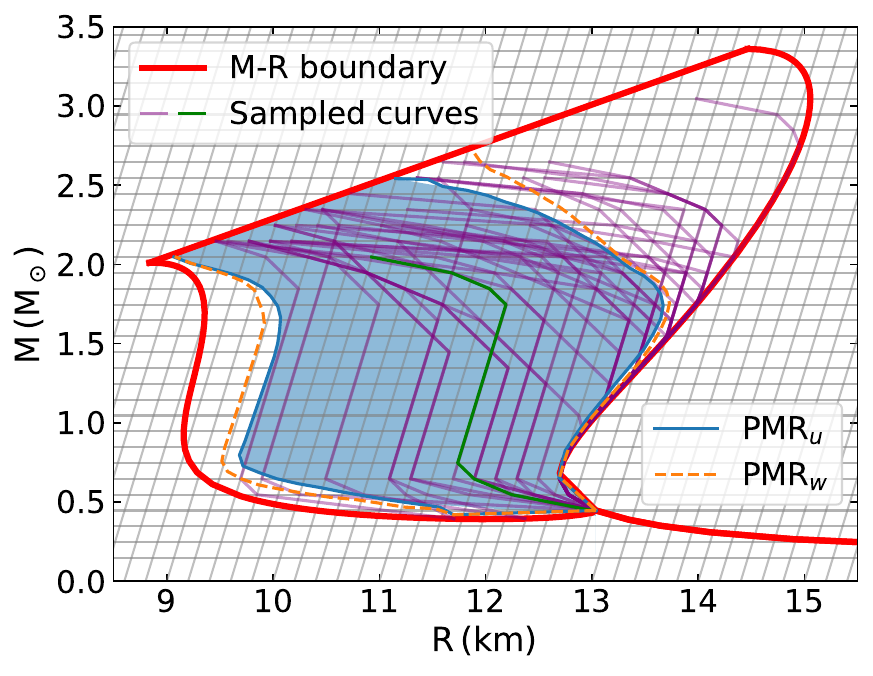}  
    \includegraphics[width=0.495\linewidth]{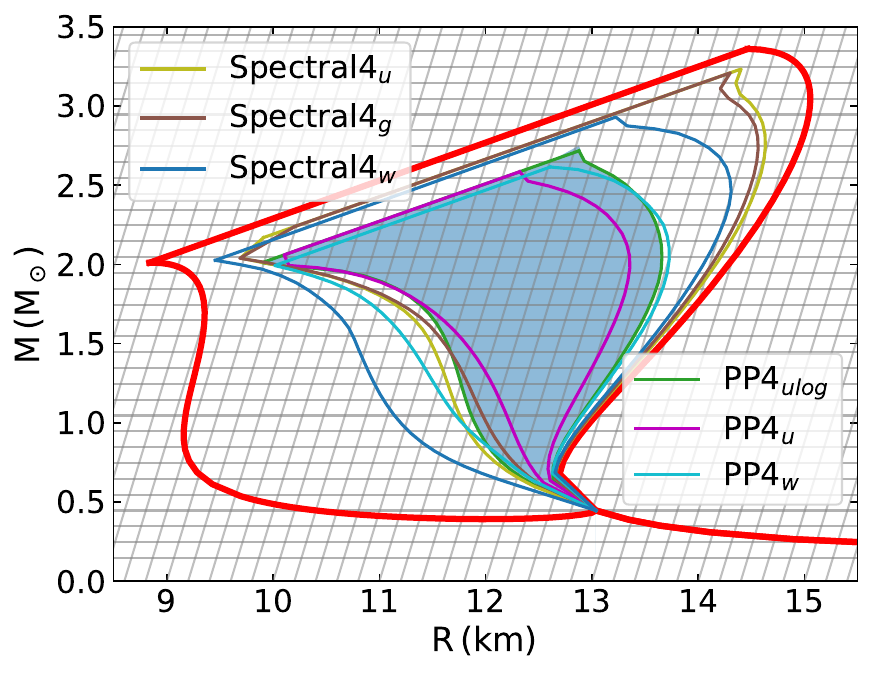}
    \caption{Left panel: An illustration of the PMR method, showing the adopted $M$-$R$ mesh (thin lines) as well as the allowed $M$-$R$ boundary (thick red line). Some randomly sampled $M$-$R$ curves are displayed as purple lines, while the green curve corresponds to the specific example defined by Eq.~\ref{eq:sample_sequence} (See Appendix \ref{ap:PMR}). 95\% confidence regions for the PMR$_u$ and PMR$_w$ priors are shown as the blue region and enclosed by the orange dashed curve, respectively.  Right panel: Together with the allowed $M$-$R$ bounds and the PMR mesh as a backdrop, the 95\% confidence regions for the traditional Bayesian methods using the PP4 and spectral parameterized EOSs are shown. For both, parameters were sampled either uniformly (subscript ``u”), log-uniformly (subscript ``ulog”), a Gaussian distribution (subscript ``g”), or through the weighting scheme for $R_{1.4}$ values (subscript ``w”).}
    \label{fig:pmr}
\end{figure*}

To generate the smallest radius for a given mass greater than $M_t$, a first-order phase transition with constant pressure $P_t$ is assumed, extending up to a critical energy density $\mathcal{E}_c\simeq 0.63\,\rm GeV\,fm^{-3}$.  Above $\mathcal{E}_c$, the EOS is taken to be causal with the sound speed $c_s=c$. $\mathcal{E}_c$ is set by the requirement that $M_{\rm max}=2.0M_\odot$ can be achieved.  On the other hand, the largest NS radius for $M>M_t$ is obtained using an EOS with $c=c_s$ for $n\ge n_t$.  The maximum mass point along the maximum radius line connects to the corresponding minimum radius point at $2.0M_\odot$, closing the allowed region.  The resulting bounded region in $M$-$R$ space is illustrated in Fig. \ref{fig:pmr} using the central value of the N$^3$LO EOS.

\subsection{Constructing the $M$-$R$ Mesh}
We proceed to parameterize the bounded $M$–$R$ space directly by constructing a discrete mesh with evenly-spaced horizontal grid lines separated by $\Delta M=0.1M_\odot$, and angled grid lines separated by $\Delta R=0.2$ km to specify $R$, as illustrated in Fig. \ref{fig:pmr}. Numerical convergence with respect to mesh spacing is demonstrated in Appendix \ref{ap:convergence}. The grid is anchored to the fixed point $(M_t,R_t)$ determined the central value together with the 1-$\sigma$ uncertainty of the N$^3$LO EOS.  Instead of using rectangular grid, we chose parallelogramed grid with a shear value $0.455 ~M_\odot^{-1}$km. This number defines the maximum $dR/dM$ allowed for the sampled M-R curves except for the part along the largest radius boundary; see Appendix \ref{ap:PMR} for details.

\subsection{Generating $M$-$R$ Curves}
Next, all possible, physically realistic, continuous $M$–$R$ curves are constructed by successively connecting grid points starting from the fixed point $(M_t,R_t)$ and ending at a point having a mass greater than or equal to the assumed minimum value of $M_{\rm max}$, namely $2.0M_\odot$.  Succeeding segments increase the mass by $\Delta M$ and can initially either follow the minimum or maximum radius boundaries, or extend to the same or smaller radial grid lines. For $M$–$R$ curves originating from the maximum-radius boundary, $dR/dM$ is required to be non-increasing along the remainder of the curve. For curves originating from the minimum-radius boundary or directly from $(M_t,R_t)$, the inverse slope $dR/dM$ is initially allowed to be non-decreasing; after an arbitrary transition point, this condition is replaced by a non-increasing $dR/dM$.
This criterion allows $dR/dM$ to change sign and therefore permits backward-bending of an $M$–$R$ curve within the range of slopes.
The curve can instead terminate if the mass is equal to or larger than the minimum assumed $M_{\rm max}$. We do not consider the possibility to extend to a new point with the same mass and a smaller radius, as first-order phase transitions are not well fit by our inversion formulae, and extension to a larger radius is disallowed by requirements of causality or thermodynamic stability (as determined below). 
Segments that would extend to a mesh point outside of the $M$-$R$ boundary are disallowed.  At present, each segment extends only to the next larger mass, but in the future we will allow segments to span more than one mass unit.

Each $M$-$R$ curve is then accepted or rejected depending on whether or not its inferred $P$-${\mathcal E}$ relation determined by Eq.~(\ref{eq:eqfit1}) satisfies causality and thermodynamic stability at every density. In this way, $M$-$R$ space is filled with the surviving $M$-$R$ curves rather than with the traditional series of $M$-$R$ curves generated from a parameterized EOS.  

The current implementation of the PMR method has several limitations. First, because the radial grid has a finite slope, the largest positive inverse slope $dR/dM$ that can be represented is limited by the shear of the parallelogrammed grid. Strongly backward-bending segments outside this range are therefore excluded by the present discretization. This is a compromise because the largest value of $dR/dM$ at arbitrary points is determined by the causality, which is unable to be obtained with the current inversion scheme \citep{Sun2025}. 
Second, because the mesh is relatively coarse, linear interpolation produces $M$–$R$ curves with noticeable kinks. However, 
the convergence study in Appendix \ref{ap:convergence} shows that the principal uncertainty measures used in this work are insensitive to the adopted grid spacing. Exact smoothness at every point, however, is not imposed. Third, the present inversion scheme is not designed for disconnected or more complicated $M$-$R$ topologies produced by sufficiently strong phase transitions, including twin-star branches. Such configurations are therefore not included in the present analysis.

The probability $\mathcal{P}$ of a given $M$-$R$ curve $\mathcal{M}_j(M,R)$ under this framework is the posterior probability distribution computed using Bayes' theorem
\[
    {\cal P}(\mathcal{M}_j(M,R)|{\cal D})\propto{\Pi_i{\cal P}({\cal D}_i|\mathcal{M}_j(M,R)){\cal P}(\mathcal{M}_j(M,R))},
\]
where $i\in{1,2,3}$ corresponds to data from the three NICER observations PSR J0740+6620 \citep{Salmi2024}, PSR J0437-4715 \citep{Choudhury2024} and PSR J0030+0451 \citep{Riley_2019}. Note that all points along the same $M$-$R$ curve share an identical posterior probability $\mathcal{P} = \mathcal{P}(\mathcal{M}_j(M,R)|\mathcal{D})$.
Once the posterior probability distribution of $M$-$R$ curves is obtained, we derive the corresponding pressure-energy density ($P$-$\mathcal{E}$) posterior using Eq.~(\ref{eq:eqfit1}).

One distinction between the PMR method and traditional Bayesian approaches lies in the fitting formula uncertainty. Although the fitting errors are typically below 1\%, they should be taken into account. Specifically, we assume a Gaussian-like error distribution, so the corrected $P$-${\mathcal E}$ probability density for a $M$-$R$ curve can be expressed as
\begin{eqnarray}
    \hspace{-1em} f(\mathcal{E}, P) &=& \sum_i\frac{\mathcal{P}_i}{2\pi\sigma_{P_i}\sigma_{\mathcal{E}_i}}\exp\left[-\frac{(\mathcal{E}-\mu_{\mathcal{E}_i})^2}{2\sigma_{\mathcal{E}_i}^2}-\frac{(P-\mu_{P_i})^2}{2\sigma_{P_i}^2}\right],\nonumber\\
    &&~\label{eq:eqfit_error}
\end{eqnarray}
where $i$ denotes a fractional $M_{\rm max}$  point on that curve, $\mu_{\mathcal{E}_i}$ and $\mu_{P_i}$ are the central values computed from the analytic inversion formulae, $\sigma_{\mathcal{E}i}$ and $\sigma_{P_i}$ are the given mean uncertainties of those values  \citep{Sun2025},  and $\mathcal{P}_i$ is the posterior probability associated with that point. The total probability density distribution, $f_{\text{tot}}(\mathcal{E},P)$, is then obtained by summing over all $M$–$R$ curves sampled in the Bayesian analysis, from which the EOS distribution can be directly derived by integrating over the $M$–$R$ space. 
Resulting confidence bands in $P$-${\mathcal E}$ space are  shown in the left panel of Fig. \ref{fig:cp1}.

Note the initial probabilities of each $M$-$R$ curve (before introduction of observational data) are equal, but the probabilities of a given grid point $(M_i,R_j)$, determined from
the number of times $M$-$R$ curves pass through that point, are not.  The initial grid point probabilities effectively form a prior $M$-$R$ space probability distribution.

\subsection{Choices of $M$-$R$-space Priors \label{sec:pmr_priors}}
We systematically explore three representative prior distributions to illustrate the impact of prior assumptions on the inferred EOS posteriors in the PMR method.

\textbf{Prior 1, Uniform sampling in $M$–$R$ Space:} Each $M$–$R$ curve is granted equal weight. The resulting density distribution of grid points is non-uniform, as more curves populate the central region than the boundaries. This is expected due to the central limit theorem (curves near the boundaries require more extreme physical conditions).  The 95\% confidence region for this  prior, labeled PMR$_u$, is shown as the blue shaded region in the left panel of Fig. \ref{fig:pmr}. 
  The count at any grid point naturally decreases with increasing distance from the fixed point $(M_t,R_t)$.

\textbf{Prior 2, Uniform Distribution of $R_{1.4}$ Values:} To achieve a more radially uniform grid density, we first group $M$-$R$ curves by their $M_{\rm max}$ value.  We could just treat each group equally and obtain a more uniform radius distribution by weighting curves by the inverse count at a typical NS radius, $R_{1.4}$.  However, the curves having large $M_{\rm max}$ cannot reach small $R_{1.4}$ values.  Instead, we correct by additionally weighting them by the number of curves within each $M_{max}$ group. This yields a more uniform probability distribution of $R_{1.4}$ values for each $M_{\rm max}$ group. To prevent a small number of sparsely sampled models from dominating the ensemble, the inverse-density weights are capped at a maximum factor of 100 relative to the minimum assigned weight before normalization. We verified that increasing this cap to 500 or 1000 changes the resulting 95\% confidence regions only slightly. Its 95\% confidence region, labeled PMR$_w$, is enclosed by the magenta dot-dashed curve in the left panel of Fig. \ref{fig:pmr}.

\begin{figure*}
\includegraphics[width=0.495\linewidth]{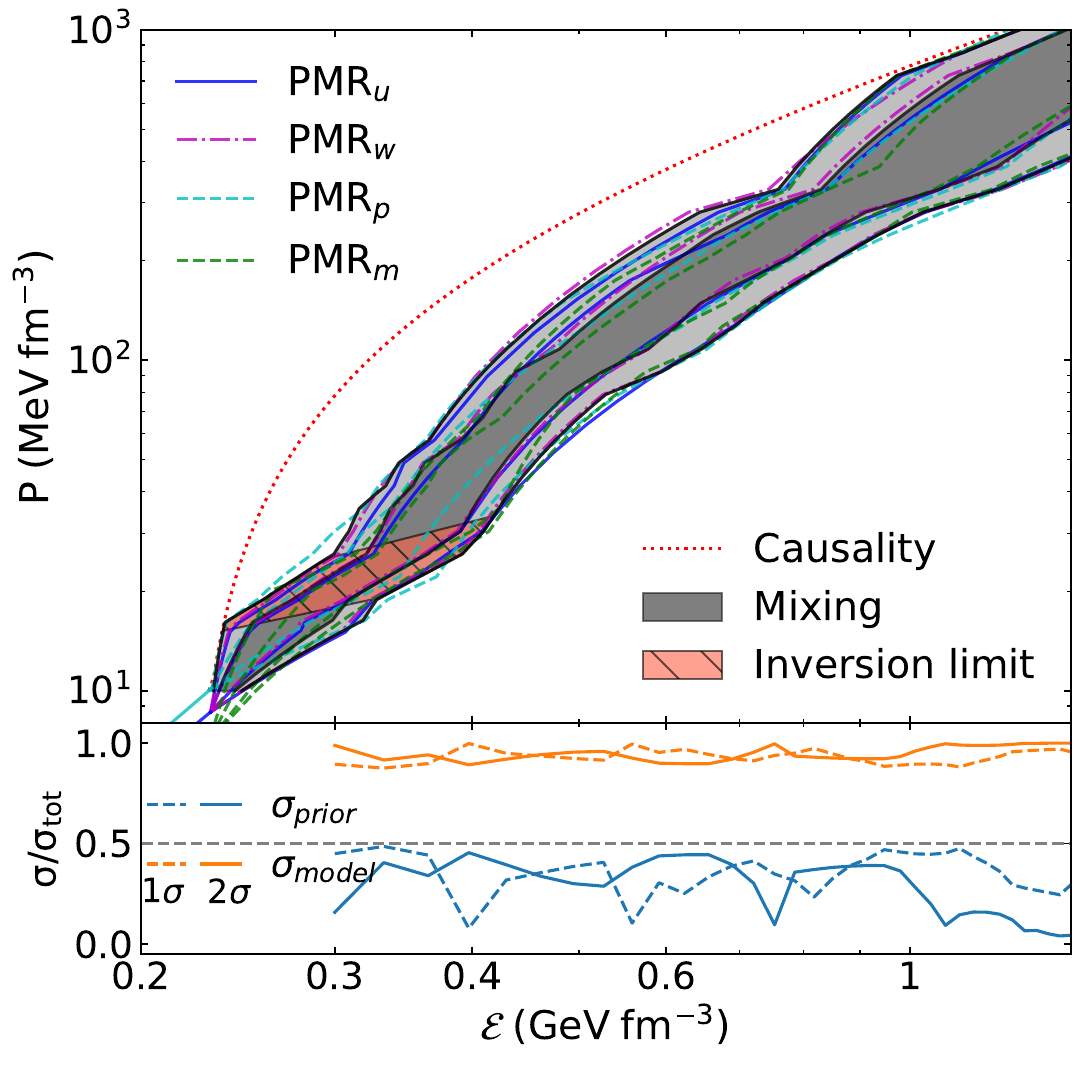}
\includegraphics[width=0.495\linewidth]{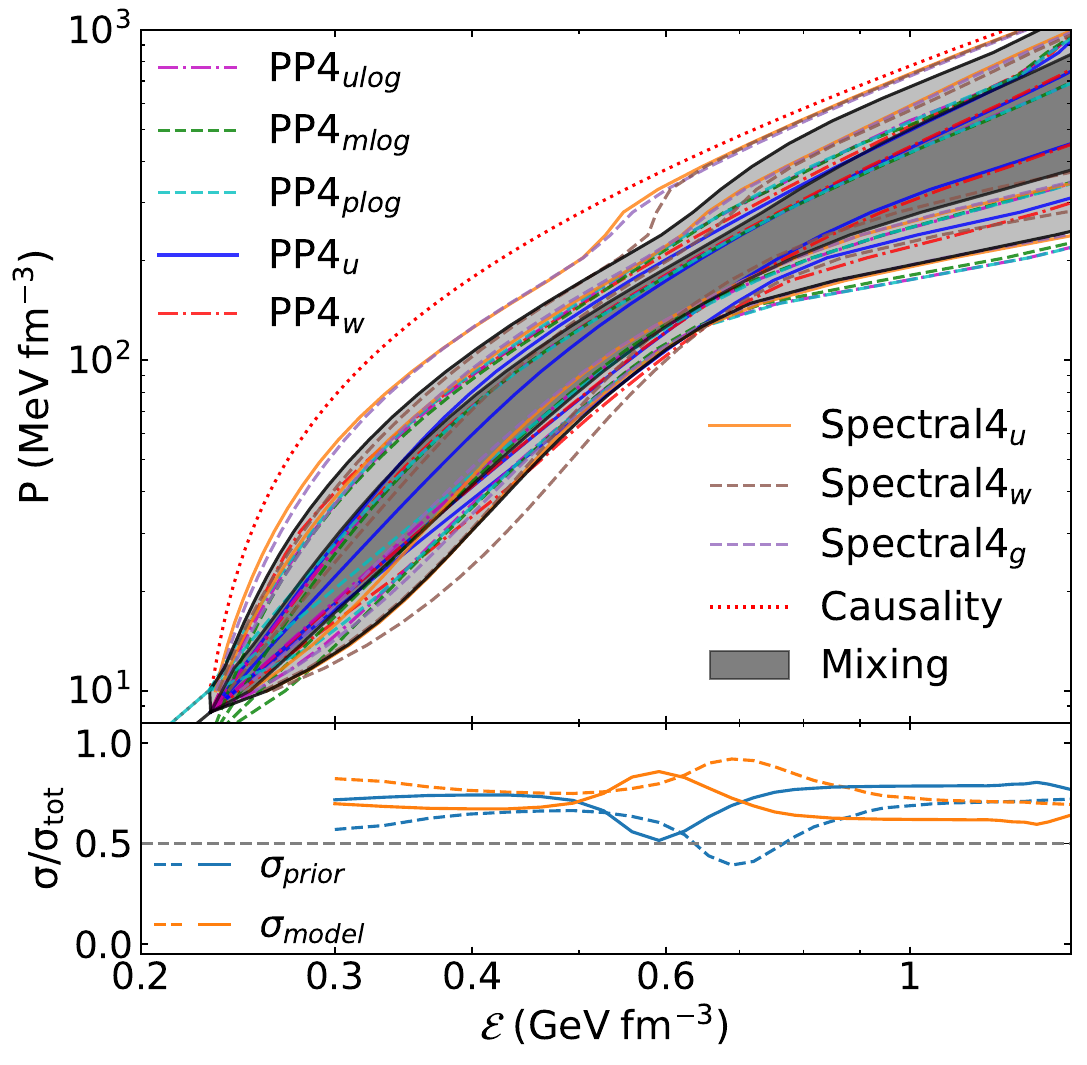}
    \caption{Comparison of Bayesian results of PMR (left) and traditional methods (right). Left: the upper panel shows the 1-$\sigma$ and 2-$\sigma$ boundaries of three priors discussed in \emph{Methodology}. The red dotted line indicates the causality constraint, and the gray shaded region refers to the model mixing of all priors, discussed in \emph{Results}. The coral-colored shaded region is a region our inversion formula may not reach, corresponding to masses less than $1/3M_{\rm max}$, below which the boundaries are obtained by linear interpolation. The first lower panel shows the ratios of $\sigma_{prior}$ and $\sigma_{model}$ to $\sigma_{tot}$ for $300-1400 $ MeV fm$^{-3}$ for 1-$\sigma$ (dashed lines) and 2-$\sigma$ (solid lines), respectively.   Right: similar to the left panel, but for traditional Bayesian methods.
    \label{fig:cp1}}
\end{figure*}

\textbf{Prior 3, Variation of the N$^3$LO Boundaries:} To straightforwardly assess the impact of systematic uncertainties associated with the N$^3$LO EOS, we vary it between its stated 1-$\sigma$ upper and lower bounds, while otherwise uniformly sampling EOS curves as in Prior 1. These prior choices are labeled in Fig. \ref{fig:cp1} as PMR$_p$ and PMR$_m$, standing for plus and minus, respectively, and change the fixed point to which the mesh is anchored slightly. For PMR$_p$, $(M_t,R_t) \simeq (0.52\,M_\odot, 13.2$ km), whereas for PMR$_m$, $(M_t,R_t) \simeq (0.35\,M_\odot, 13.0$ km). 

\subsection{Benchmark with Traditional Bayesian Inferences \label{sec:trad_priors}} 
To benchmark our PMR framework, we also undertake traditional Bayesian calculations using two commonly used EOS parameterizations: a four-segment piecewise-polytropic model (PP4) and a four-parameter spectral representation (Spectral4), described in Appendix \ref{ap:PP4} and \ref{ap:Spectral4}, respectively. We examine linearly uniform parameter sampling for both PP4 and Spectral4 (PP4$_u$ and Spectral4$u$), logarithmically uniform sampling for PP4 (PP4${ulog}$), and Gaussian sampling for Spectral4 (Spectral4$_g$). We also introduce the same weighting scheme as PMR$_w$ to get uniform distribution of $R_{1.4}$ values for linearly uniform sampling of both models (labeled PP4$_w$ and Spectral4$_w$). Furthermore, we vary as well the N$^3$LO EOS, using the same notations as for the PMR approach, to ensure a consistent comparison. A minimum $M_{\rm max}$ value of $2.0M_\odot$ is again adopted, and causality and thermodynamic stability are enforced in all cases. 

The right-hand panel of Fig. \ref{fig:pmr} shows the corresponding 95\% confidence regions for traditional Bayesian approaches with various assumed priors. It is apparent that the small radius region is under-represented compared to the PMR method, irrespective of  assumed priors in either method.  On the other hand, the PMR method seems to under-populate the large radius and high mass regions. It is also clear that the PP4 parameterization  is not able to fully represent the region near the upper boundary of the allowed $M$-$R$ space. A quantitative comparison is provided in Table \ref{tab:summary}, which shows that the PMR priors considered here generally achieve larger coverage fractions than any of the selected traditional priors. Overall, the differences between uniform, log-uniform and Gaussian parameter sampling are relatively small.
\begin{table*}[!ht]
\centering

    \begin{tabular}{c|c|c|c|c|cc|cc}
    \Xhline{1pt}
    \multirow{3}{*}{Priors} & \multirow{3}{*}{Sampling}& \multirow{3}{*}{Weighting} & \multirow{3}{*}{N$^3$LO choice} & \multirow{3}{*}{Coverage fraction} &  \multicolumn{2}{c|}{$\mu(\sigma_{\rm prior})$} & \multicolumn{2}{c}{$\mu(\sigma_{\rm tot})$} \\
     & & & & & \multicolumn{2}{c|}{$\rm [MeV\;fm^{-3}]$}& \multicolumn{2}{c}{$\rm [MeV\;fm^{-3}]$} \\
     \cline{6-9}
      & & & & & $1\sigma$ & $2\sigma$ & $1\sigma$ & $2\sigma$\\
     \Xhline{1pt}
     PMR$_u$ & \multirow{4}{*}{Uniform} & Equal & \multirow{2}{*}{Central} & 0.641 & \multirow{4}{*}{72.08} & \multirow{4}{*}{66.97}& \multirow{4}{*}{167.6}& \multirow{4}{*}{283.5}\\ 
     PMR$_w$ & & $R_{1.4}$-weighted & & 0.702 & & & &\\
     \cline{3-4}
     PMR$_p$ & & \multirow{2}{*}{Equal} & Upper (1$\sigma$)& 0.655 & & & &\\
     PMR$_m$ & & &Lower (1$\sigma$) & 0.652 & & & &\\ 
     \hline
     PP4$_{ulog}$ & \multirow{3}{*}{Log-uniform} & \multirow{3}{*}{Equal} & Central & 0.387 & \multirow{8}{*}{124.9} & \multirow{8}{*}{286.6}& \multirow{8}{*}{185.3} & \multirow{8}{*}{372.2}\\
     PP4$_{plog}$ & & & Upper (1$\sigma$) & 0.367 &  & &  & \\
     PP4$_{mlog}$ & & &Lower (1$\sigma$)& 0.358 &  & &  & \\
     \cline{2-4}
     PP4$_{u}$ & \multirow{2}{*}{Uniform} & Equal&\multirow{2}{*}{Central} & 0.289 & & &  &\\
     PP4$_{w}$ & & $R_{1.4}$-weighted & & 0.408 & & &  &\\
     \cline{1-5}
     Spectral4$_{u}$ & \multirow{2}{*}{Uniform} & Equal & \multirow{3}{*}{Central} & 0.617 &  & &  &\\
     Spectral4$_{w}$ & & $R_{1.4}$-weighted & & 0.601 &  & &  &\\
     \cline{2-2}
     Spectral4$_{g}$ & Gaussian & Equal & & 0.581 &  & &  &\\
     \Xhline{1pt}
    \end{tabular}
    \caption{Summary of the differences among all selected priors. The first three columns indicate the sampling prescription, weighting prescription, and choice of N$^3$LO boundaries, respectively, discussed in Sec.~\ref{sec:pmr_priors} and \ref{sec:trad_priors}. Coverage fraction is the area enclosed by 95\% confidence region of each prior (for example, the blue shaded regions in Fig.~\ref{fig:pmr}) divided by the total area of the physically allowed $M$-$R$ regions enclosed by the red boundaries in Fig.~\ref{fig:pmr}. The last two columns show the mean values of $\sigma_{\rm prior}$ and $\sigma_{\rm tot}$ for the PMR and traditional methods, respectively, for both 1$\sigma$ and 2$\sigma$ bounds. \label{tab:summary}}
\end{table*}

\section{Results \label{sec:results}}
We highlight two important issues that arise from our analysis.   First, we examine the relative importance of prior modeling uncertainties in the PMR and traditional EOS Bayesian methods relative to the observational uncertainties.  Second, we examine the relative accuracies of the two approaches when confronted with the same astrophysical data. 

To determine the systematic prior uncertainties of each inference scheme, we applied the {\it model mixing method}, a statistical framework widely used for combining model ensembles in nuclear and astrophysical analyses (e.g., \cite{PhysRevC.111.035804, PhysRevD.99.084049}). At each energy density $\mathcal{E}$, the posterior samples obtained from each prior choice define a distribution of possible pressures $P$. We represent this distribution by a kernel density estimate (KDE), denoted by $\hat f_i(P\,|\,\mathcal{E})$, where $i$ labels the prior choice.

Assigning equal weight to each of the $N$ priors, the mixed KDE is
\begin{eqnarray}
    \hat f_{\rm mix}(P\,|\,\mathcal{E})
    &=&
    \frac{1}{N}\sum_{i=1}^{N}\hat f_i(P\,|\,\mathcal{E}) .
\end{eqnarray}
The corresponding mean pressure is
\begin{eqnarray}
    \bar P(\mathcal{E})
    &=&
    \frac{1}{N}\sum_{i=1}^{N}\bar P_i(\mathcal{E}),
\end{eqnarray}
where
\begin{eqnarray}
    \bar P_i(\mathcal{E})
    =
    \int P\,\hat f_i(P\,|\,\mathcal{E})\,dP .
\end{eqnarray}
The total variance of the pressure at fixed $\mathcal{E}$ can then be decomposed into two contributions:
\begin{eqnarray}
    \sigma_{\rm tot}^2(\mathcal{E})
    &=&
    \int
    \left[
    P-\bar P(\mathcal{E})
    \right]^2
    \hat f_{\rm mix}(P\,|\,\mathcal{E})\,dP
    \label{eq:sig_tot}
    \nonumber\\
    &=&
    \sigma_{\rm model}^2(\mathcal{E})
    +
    \sigma_{\rm prior}^2(\mathcal{E}) .
    \label{eq:sig_tot}
\end{eqnarray}
Here, $\sigma_{\rm model}$ denotes the average intrinsic uncertainty within individual prior choices,
\begin{eqnarray}
    \sigma_{\rm model}^2(\mathcal{E})
    &=&
    \frac{1}{N}\sum_{i=1}^{N}\sigma_i^2(\mathcal{E})
    \nonumber\\
    &=&
    \frac{1}{N}\sum_{i=1}^{N}
    \int
    \left[
    P-\bar P_i(\mathcal{E})
    \right]^2
    \hat f_i(P\,|\,\mathcal{E})\,dP ,
\label{eq:sig_model}\end{eqnarray}
whereas $\sigma_{\rm prior}$ quantifies the spread of the mean pressures induced by different prior choices,
\begin{eqnarray}
    \sigma_{\rm prior}^2(\mathcal{E})
    &=&
    \frac{1}{N}\sum_{i=1}^{N}
    \left[
    \bar P_i(\mathcal{E})-\bar P(\mathcal{E})
    \right]^2 .
\label{eq:sig_prior}\end{eqnarray}
A comparison of these two contributions to uncertainties, as well as their totals, employing a variety of assumed prior distributions in the PMR and traditional Bayesian approaches, are shown in the lower panels of Fig.~\ref{fig:cp1}.

Overall, the posteriors of PMR exhibit behavior broadly similar to those from traditional methods across the full energy-density range, although the PMR confidence boundaries are somewhat less smooth at low energy density. This behavior is associated with the current inversion scheme as the fitting uncertainty increases in this regime \citep{Sun2025}. It is expected that improved inversion formulae for lower-mass configurations should reduce this effect in the future. However, the sensitivity to prior assumptions differs significantly. Quantitatively, the fraction of posterior variance attributable to prior choices, quantified by $\sigma_{prior}/\sigma_{tot}$ from Eqs.~(\ref{eq:sig_tot}) and (\ref{eq:sig_prior}), remains below 50\% in the PMR method, while traditional frameworks show substantially higher sensitivity, with mean values around 70\% and typical variations of 50–80\%. This indicates that the new framework is less affected by prior selection.

Since the density dependence of $\sigma_{prior}$, $\sigma_{model}$ and $\sigma_{tot}$ is weak, we characterize each uncertainty by the density averaged mean values of $\sigma$,

\begin{equation}
    \mu (\sigma) = \frac{1}{\mathcal{E}_{max}-\mathcal{E}_{min}} \int_{\mathcal{E}_{min}}^{\mathcal{E}_{max}} \sigma(\mathcal{E}) d\mathcal{E}, \label{eq:average_uncertainty}
\end{equation}
where $\mathcal{E}_{min}=300$ MeV fm$^{-3}$, $\mathcal{E}_{max}=1400$ MeV fm$^{-3}$.

As shown in the last two columns of Table \ref{tab:summary},the density averaged mean values of both 1-$\sigma$ and 2-$\sigma$ bounds of $\sigma_{tot}$ as well as $\sigma_{prior}$ of PMR are smaller than in the traditional approaches, indicating a moderate improvement for the new approach.  While encouraging, it should be emphasized, however, that this improvement might not apply for a broader sampling of EOS-prior-based Bayesian approaches, or for a more extensive sampling of prior assumptions.

\section{Conclusion and Outlook}
Our result show that Bayesian inference on the NS EOS is significantly influenced by prior choices which introduce sizable systematic uncertainties, therefore highlighting the importance of assessing multiple prior choices when applying Bayesian methods to NS observations.
When this is impractical, a conservative estimate of the prior-induced variance, $\sigma_{\mathrm{prior}}^2 \simeq0.5\sigma_{tot}^2$, may be adopted based on our results for typical traditional frameworks.

The newly developed PMR method offers several advantages: (1) both its priors and observational data are defined in the $M$-$R$ space, making it more direct and intuitive when comparing with observations; (2) computations are generally much faster because EOS parameterization and repeated TOV calculations are avoided; (3) its priors tend to be more uniformly distributed and more easily cover the entire allowed $M$–$R$ region, especially the small radius region as demonstrated in Fig. \ref{fig:pmr}; (4) its posteriors exhibit significantly lower sensitivity to prior assumptions; and (5) it also generally produces a smaller total uncertainty when confronted with the same observational data, at least for the limited comparisons included here. 

The method can be further improved by reducing the uncertainty of the inversion formulae from  \citet{Sun2025}.  The transformation from the $M$-$R$ relation to the $P$-${\mathcal E}$ relation in Eq.~(\ref{eq:eqfit1}) could, for example,  be modeled using a neural network \cite{fujimoto2020mapping,morawski2020neural,soma2023reconstructing,patra2025inferring,Zhou2024}, while the $M$-$R$ prior is kept unchanged and defined in the same parameter space as the $M$-$R$ likelihood. This approach can potentially incorporate a wider variety of EOSs with different degrees of freedom and achieve higher fidelity.

More broadly, our findings underscore the need for alternative inference frameworks that offer broader coverage of prior space and reduced dependence on subjective prior assumptions. Such developments will help improve both the reliability and the precision of neutron star EOS constraints from observations. Future work will explore more complex $M$–$R$ relations, including those with multiple phase transitions, enabled by improved inversion techniques capable of mapping such $M$–$R$ curves to corresponding EOSs with high accuracy. Recent progress using gradient descent initialized by the current inversion formula following \citet{Ronghao2025} has yielded encouraging results, suggesting that combining it with machine-learning approaches is a promising next step.

\begin{acknowledgments}
B.S. and J.M.L. acknowledge funding from the US Department of Energy under Grant DE-FG02-87ER40317. T.Z. acknowledges support by the Network for Neutrinos, Nuclear Astrophysics and Symmetries (N3AS) through the National Science Foundation Physics Frontier Center, Grant No. PHY-2020275. We thank Chun Huang, Alan Calder, Douglas Swesty, Sophia Han, Zidu Lin, Shuzhe Shi and Xihaier Luo for helpful discussions.
\end{acknowledgments}

\appendix
\section{Parameterized Mass-Radius (PMR) Representation} \label{ap:PMR}
There is no unique way to parameterize $M$-$R$ curves. We sample the space of allowed $M$-$R$ relations using a sequence of linear segments with uniform mass spacing and discretized radius spacing, subject to the boundaries discussed in Sec.~\ref{sec:MR_boundary}. Linear interpolation is used between adjacent grid points. The resulting mass--radius curves are written as
\begin{eqnarray}
R(M) &=& R_i + \left[\frac{R_{i+1}-R_i}{M_{i+1}-M_i}\right] (M-M_i), ~~~~~M_i<M<M_{i+1}
\end{eqnarray}
where
\begin{eqnarray}
R_{i+1}&=&R_i+N_i\Delta R+ \left[\frac{dR}{dM}\right]_{\rm max} \Delta M, \\
M_{i+1}&=&M_i+\Delta M.
\end{eqnarray}
We take $\Delta M=0.1\,M_\odot$, $\Delta R=0.2$ km, and
\begin{equation}
\left[\frac{dR}{dM}\right]_{\rm max}
=0.455 ~\frac{\textrm{km}}{M\odot}.
\end{equation} Each $N_i\in\mathbb{Z}_{\leq 0}$ can take any non-positive integer value, while the sequence ${N_i}$ is required to be \emph{unimodal}. For example, for an $M$-$R$ curve starting from $
(M_{i=0},R_{i=0})=(M_t,R_t)
=(0.45\,M_\odot,13.0~{\rm km})$
the sample sequence
\begin{equation}
\{N_i\}=
\{-4,-2,-1,0,0,0,0,0,0,0,0,0,0,-1,-2,-4\}
\label{eq:sample_sequence}
\end{equation}
generates an $M$-$R$ curve that terminates at $M=2.05\,M_\odot$ and has a radius of $12$ km at $M=1.4\,M_\odot$, corresponding to the green curve in the left panel of Fig.~\ref{fig:pmr}.

The minimum and maximum radius boundaries in Sec.~\ref{sec:MR_boundary} can likewise be discretized and represented by the arrays $\{B_i^-\}$ and $\{B_i^+\}$, respectively, using the same functional form as above, except that each $B_i^\pm\in\mathbb{Q}$ may take a rational value. Because a sampled $M$-$R$ curve is allowed to originate from either the minimum or maximum radius boundary, an initial portion of either boundary sequence, $\{B_i^-\}$ or $\{B_i^+\}$, may be concatenated with an integer sequence $\{N_i\}$. At the transition from the boundary to the radius grid, the final element of $\{B_i^-\}$ is rounded up to the nearest radial grid line, whereas the final element of $\{B_i^+\}$ is rounded down. The resulting concatenated sequence is also required to be \emph{unimodal}.

\setcounter{equation}{0}
\section{Convergence Test \label{ap:convergence}}
To assess numerical convergence of the PMR framework, we repeated the entire analysis using three different horizontal mesh spacings, $\Delta M=0.10$, $0.15$, and $0.20\,M_\odot$, while keeping all other components of the algorithm unchanged. We verify that the results of the PMR method are insensitive to the adopted mass spacing. The quantities reported in Table~\ref{table:1} are the density averaged mean values of the uncertainty ratios, $\mu(\sigma_{\rm prior}^{\rm PMR}/\sigma_{\rm tot}^{\rm PMR})$, $\mu(\sigma_{\rm prior}^{\rm PMR}/\sigma_{\rm prior}^{\rm trad})$, and $\mu(\sigma^{\rm PMR}_{\rm tot}/\sigma^{\rm trad}_{\rm tot})$ defined as
\begin{equation}
    \mu (\frac{\sigma_i}{\sigma_j}) = \frac{1}{\mathcal{E}_{max}-\mathcal{E}_{min}} \int_{\mathcal{E}_{min}}^{\mathcal{E}_{max}} \frac{\sigma_i(\mathcal{E})}{\sigma_j(\mathcal{E})} d\mathcal{E},
\end{equation}
similar to Eq.~(\ref{eq:average_uncertainty}) in Sec.~\ref{sec:results}. In particular, the average contribution of prior uncertainty, quantified by $\mu(\sigma_{\rm prior}^{\rm PMR}/\sigma_{\rm tot}^{\rm PMR})$, changes by less than $0.03$. The average ratios of the prior uncertainties and total uncertainties between the PMR and traditional Bayesian approaches vary by less than $0.02$ and $0.01$, respectively. These small differences indicate that the conclusion that the PMR framework is less sensitive to prior choices than traditional methods is robust against the mesh resolution, and that $\Delta m=0.1\,M_\odot$ is sufficient for practical applications.
\begin{table}[h]
\centering
    \begin{tabular}{c|cc|cc|cc}
    \hline
    \multirow{2}{*}{$\Delta m$} & \multicolumn{2}{c|}{$\mu(\sigma_{\rm prior}^{\rm PMR}/\sigma_{\rm tot}^{\rm PMR})$} &  \multicolumn{2}{c|}{$\mu(\sigma^{\rm PMR}_{\rm prior}/\sigma^{\rm trad}_{\rm prior})$}&  \multicolumn{2}{c}{$\mu(\sigma^{\rm PMR}_{\rm tot}/\sigma^{\rm trad}_{\rm tot})$}  \\
     \cline{2-7}
      & $1\sigma$ & $2\sigma$ & $1\sigma$ & $2\sigma$ & $1\sigma$ & $2\sigma$ \\
     \hline
     0.1 & 0.352 & 0.267 & 0.422& 0.246 & 0.758 & 0.682 \\ 
     0.15 & 0.362 & 0.283 & 0.435& 0.259 & 0.760 & 0.678\\
     0.2 & 0.356 & 0.291 & 0.424& 0.265 & 0.753 & 0.674\\
     \hline
    \end{tabular}
    \caption{Convergence test for different horizontal spacing ($\Delta m$). \label{table:1}}
\end{table}

\section{Piecewise Polytropic (PP) Representation \label{ap:PP4}}
\setcounter{equation}{0}

Following the framework proposed by \citet{read2009constraints}, high-density, cold nuclear equations of state (EOSs) can be well approximated by a sequence of piecewise-polytropic segments. Each segment is described by the polytropic equation of state $P=K_in^{\gamma_i}$ for the region $n_{i-1}<n<n_i$, where $P$ is the pressure, $n$ is the baryon number density, and the constants $K_i$ and adiabatic indices $\gamma_i$ refer to the $i$-th segment. Matching conditions at the core-crust transition, together with the continuity of $P$ and the energy density ${\mathcal E}$ at the boundaries, determine $K_i$, leaving 2 free parameters for each segment—namely, $\{n_i, \gamma_i\}$, or, equivalently, $\{n_i,\,P_i\}$. Unlike earlier implementations in which the transition densities $n_i$ were fixed \citep{zhao2018tidal}, we treat the segment boundaries $n_i$ as additional free parameters. In this work, we adopt a modified four-segment PP4 representation to increase flexibility at high densities, corresponding to $i=1, 2, 3, 4$. Within segment i, the energy density is given by
\begin{equation}
{\mathcal E}={\mathcal E}_{i-1}{n\over n_{i-1}}+{P-P_{i-1}(n/n_{i-1})\over\gamma_i-1},\quad n_{i-1}\le n\le n_i,
\end{equation}
which follows directly from the standard thermodynamic relation between pressure and energy density for a polytropic EOS. Consistent with PMR method, we choose the transition density $n_0 = 1.5 n_s$, rather than matching directly to a crust EOS. The corresponding energy density ${\mathcal E}_0$ and pressure $P_0$ are determined from N$^3$LO $\chi$EFT calculations \citep{Drischler_2021}. In the interval $n_0 < n < n_{\max}=12 n_s$, we select four intermediate densities and sort them as
$n_0 < n_1 < n_2 < n_3 < n_4 < n_{\max}$. The associated pressures are labeled such that
$P_0 < P_1 < P_2 < P_3 < P_4 < P_{\max}$. For the linear prior, $P_{1,2,3,4}$ are sampled uniformly between $P_0$ and $P_{\max}=3000$ MeV fm$^{-3}$. For the logarithmic prior, $\ln P_{1,2,3,4}$ are sampled uniformly between $\ln P_0$ and $\ln P_{\max}$, which places more weight on the low-pressure end. In both cases, EOS parameter sets that violate causality are excluded.

Bayesian analysis based on PP4 is subject to prior assumptions associated with the finite number of segments, taken to be four in this work, as well as the choice of the energy density and pressure at the lower baryon number density boundary $n_0$. The sampling of the transition densities $n_i$ and pressures $P_i$ is not unique and should therefore also be regarded as part of the prior specification. Additional physical constraints, such as causality and thermodynamic stability, are universal requirements for any EOS. However, their impact within the PP4 framework is implementation dependent. For example, thermodynamic stability is enforced by requiring an ordered sequence of pressures, $P_1 < P_2 < P_3 < P_4$, rather than by explicitly rejecting configurations with $P_i > P_j$ for $i < j$. Similarly, violations of causality are more likely to occur at the high density end of each segment, since the corresponding adiabatic indices \( \gamma_i \) are typically larger than 1.

\section{Spectral Representation \label{ap:Spectral4}}
\setcounter{equation}{0}

An alternative and complementary approach is the spectral representation developed by \citet{lindblom2010spectral}, which provides an efficient parametrization of smooth, realistic EOSs with relatively few parameters. The method expands the logarithm of the adiabatic index $\Gamma$ as a Taylor series in the logarithm of pressure:
\begin{eqnarray}
\textrm{ln}\,\Gamma(x)=\sum_{i=0}^i \gamma_i x^i, \quad
x=\textrm{ln}\frac{P}{P_0},
\end{eqnarray}
where $P_0$ is the pressure at a matching density $n_0$, consistent with the PP representation. The adiabatic index $\Gamma$ is related to the thermodynamic derivatives of pressure and energy density via
\begin{eqnarray}
\Gamma(x) = \frac{d\,(\textrm{ln}\,P)}{d\,(\textrm{ln}\,n)} = \frac{{\mathcal E}+P}{P} \frac{d P}{d {\mathcal E}}.
\end{eqnarray}
Integrating this relation yields the energy density and baryon number density as
\begin{eqnarray}
{\mathcal E}=\frac{n}{n_0} \left[{\mathcal E}_0 + P_0 \int_0^x \frac{n_0\,e^{x'}}{n'\,\Gamma(x')} dx'\right], \quad
\textrm{ln}\,\frac{n}{n_0}=\int_{0}^x\frac{dx'}{\Gamma(x')} \label{eq:integration_baryon}.
\end{eqnarray}
In the Spectral4 parameterization adopted here, $\ln\Gamma(x)$ is expanded up to third order, including four parameters $\gamma_0$, $\gamma_1$, $\gamma_2$, $\gamma_3$. Numerical evaluation is performed using quadratic integration for number density 
and double-quadratic integration for energy density in Eq.~(\ref{eq:integration_baryon}) 
. To improve computational efficiency, interpolation is employed after the initial tabulation of the EOS. Boundary matching at the core–crust interface ensures thermodynamic consistency.
The original proposed uniform prior ranges for the spectral parameters are $\gamma_0\in[0.2,2]$, $\gamma_1\in[-1.6,1.7]$, $\gamma_2\in [-0.6,0.6]$, $\gamma_3\in [-0.02,0.02]$ \citep{abbott2018gw170817,carney2018comparing,miller2019constraining}, based on $P_0\approx 5.4\times 10^{32}$ corresponds to $n_0 = 0.5 n_s$. In this work, we choose the transition density $n_0=1.5n_s$ to be consistent with PMR and PP representations. Therefore, we modify the prior range to $\gamma_0\in[0,3]$, $\gamma_1\in[-3,3]$, $\gamma_2\in [-1,1]$, $\gamma_3\in [-0.2,0.2]$, which cover the vast majority of the faithful EOSs. We also test the Gaussian prior on these parameter which follows $\gamma_0 \in \mathcal{N}(1.5,0.75^2)$, $\gamma_1\in\mathcal{N}(0,1.5^2)$, $\gamma_2\in \mathcal{N}(0,0.5^2)$, $\gamma_3\in \mathcal{N}(0,0.1^2)$. In both cases, EOS parameter sets that violate causality and stability are excluded.

Bayesian analysis based on the Spectral4 is likewise subject to several prior assumptions. These include the choice of truncation order in the expansion of $\ln \Gamma(x)$, which is fixed to third order in this work, as well as the selection of the matching density $n_0$ and the corresponding pressure $P_0$ and energy density ${\mathcal E}_0$. The adopted prior ranges or distributions for the spectral coefficients $\{\gamma_0,\gamma_1,\gamma_2,\gamma_3\}$ constitute an additional and important source of prior dependence. The intrinsic smoothness of the spectral representation further suppresses sharp transition in the EOS, e.g. due to strong phase transitions. In addition, thermodynamic stability and causality must be verified explicitly over the entire density range, rather than being enforced through parameter choice and ordering as in the PP4 representation.

\bibliographystyle{aasjournal}
\bibliography{ref}

\end{document}